\documentstyle[aps,prl,preprint]{revtex}
\begin{document}
\bibliographystyle{unsrt}

\draft
\title{Intrinsic localized modes in the charge-transfer-solid PtCl}
\author{K. Kladko$^1$, J. Malek$^{2}$ and A. R. Bishop$^1$}
\address{$^1$ 
Theoretical Division and Center for Nonlinear Studies,
Los Alamos National Laboratory, Los Alamos,
87545 NM, USA \\
$^2$ Institute of Physics, Na Slovance 2, 18221 Prague 8, Czech Republic
}
\date{\today}
\maketitle
\begin{abstract}
We report a theoretical analysis of intrinsic localized modes in a
quasi-one-dimensional charge-transfer-solid
$[Pt(en)_2][Pt(en)_2 Cl_2](ClO_4)_4$(PtCl). 
We discuss strongly nonlinear features of  resonant Raman overtone 
scattering measurements on 
PtCl, arising from quantum intrinsic localized (multiphonon) modes (ILMs)  
and ILM-plus-phonon states.  
We show, that Raman scattering data displays  clear signs of a non-thermalization of  lattice
degrees-of-freedom, manifested in a nonequilibrium density
of  intrinsic localized modes.
Adiabatic lattice 
dynamics is used in a model two-band Peierls-Hubbard Hamiltonian, including 
a screened Coulomb interaction between neighboring sites. The Hamiltonian is diagonalized
on a finite chain.   
The calculated adiabatic potential for Peierls distortion of the $Cl$ sublattice
displays  characteristic non-analytic points, 
related to a lattice-distortion-induced charge transfer.
Possible nonadiabatic effects on ILMs are discussed. 
\end{abstract}


In this paper we discuss intrinsic localized modes, i.e. multi-quanta
bound states, \cite{swanson} in a 
halogen-bridged mixed-valence transition metal complex
$[Pt(en)_2][Pt(en)_2 Cl_2](ClO_4)_4$ (en=ethylenediamine),
subsequently denoted as $PtCl$ (see \cite{bishop} and
references therein). $PtCl$ is a representative of a family of 
$MX$-chain compounds, where $M$ stands for a transition metal
(e.g., $Pt$, $Pd$, or $Ni$) in a mixed-valence (i.e. charge-disproportionated) state and $X$ is a halogen
($Cl$, $Br$, or $I$). $PtCl$  consists of a 3-dimensional crystalline array of 
charged linear chains of alternating metal ($Pt^{3+}$) and halogen 
($Cl^{-}$) ions,
with $(en)_2$ ligands attached to the metals. There are also two
$ClO_4^{-}$ ions  per unit cell to maintain charge neutrality.
The  structure of $PtCl$ is given in  Fig.1 of \cite{batistic}.
Each $Cl$ ion has two electrons in the filled $p_z$ orbital ($z$ is 
oriented along the chain axis). $Pt$ ions have on average one electron
in each $d_z^2$ orbital. There are, therefore, three valence electrons or 
one hole per unit cell. The ground state of $PtCl$ displays  
a very strong charge density wave (CDW) structure. 
Holes are redistributed to produce
an alternating sequence of nominally $Pt^{2+}$ and $Pt^{4+}$ ions  
that have nearly zero and  two holes, respectively.
$Cl$ ions then move strongly closer to $Pt^{4+}$ atoms. As a result of this periodic lattice distortion, 
hopping matrix elements from
$Pt^{+4}$ ions increase and holes gain  energy by  virtual hops from 
$Pt^{+4}$ to $Cl$. 
This strong Peierls-distorted-disproportionation phase is well described by  $Cl-Pt^{4+}-Cl$
trimers, alternating with weakly coupled empty $Pt^{2+}$ sites: The $Pt^{+4} - Cl$ distance is 2.318 \AA, and
the $Pt^{+2} - Cl$ distance is 3.085 \AA.  
The electron-electron 
repulsion $U$ on $Pt$ sites is  not sufficient to destroy the CDW phase or to
significantly reduce  
its magnitude. Therefore, it is not necessary to explicitly include
correlations 
as far as the magnitude of the
CDW is concerned. Indeed, as shown in \cite{bishop}, one may introduce
effective free-electron model parameters that give correct values
for the ground state uniform Peierls distortion. However, calculations in \cite{bishop}, and our 
calculations below, show that it is necessary to introduce $U$ if one
wishes to  describe {\it electronic} properties. 
The optical absorption of $PtCl$ has two peaks, at approximately $2.5$ eV and   
$5.5$ eV, corresponding essentially to $Pt^{4+} -> Pt^{2+}$ and $Pt^{4+} -> Cl$
local charge transfer excitons \cite{bishop}. The width of the  $Pt^{4+} -> Pt^{2+}$
band is of order of $0.7eV$ \cite{wada}. This corresponds to a tunneling
time of about $5.9 \cdot 10^{-15}$ seconds for a hole to tunnel 
from one
$Pt^{+2}$ site to a near-neighbor one.     
Due to the large Peierls distortion, the inter-valence charge transfer 
(IVCT) gap is large, $\approx 2.5$ eV= $2.9 \cdot 10^{4}$K. 
The energy  $E_{ph}$ of the wavevector $k=0$ optical
phonon in the system is $38$ mEv = $437$ K. 
The period of a phonon oscillation is thus equal to $1.4 \cdot10^{-13}$s.

Resonant Raman scattering (RRS)  measurements on $PtCl$ were reported in 
\cite{swanson}, and the observed strong red-shift of overtones interpreted 
convincingly in terms of multiphonon bound states,
i.e., intrinsic local modes (ILMs). Below, we adopt a simple Brownian motion picture
of a RRS event, see Fig.1.
We first consider the simplest and most probable scenario in the local
(atomic) limit appropriate for $PtCl$.
In the first stage, Fig.1a, a photon is absorbed,
and a hole is transfered from 
a $Pt^{4+}$ ion to the neighboring  $Pt^{2+}$ ion, creating a pair of
neighboring $Pt^{3+}$ ions. 
Then, in the simplest situation, after some time the hole recombines, emitting
a photon. In more complicated and less probable scenaria a number of further
hops (tunneling steps) may occur before the recombination.  
Let us first consider the simplest scenario above. After a hole is transfered
to the empty $Pt$ site the $Cl$ ion between two $Pt^{3+}$ sites is no
longer in the minimum of its adiabatic energy. Therefore, it starts to move,
transferring the electronic energy to the energy of the lattice,
Fig.1b.
After some time   the                                             
hole hops back and a photon is emitted, Fig.1c. Some
energy remains in the
lattice vibration, Fig.1d. This energy corresponds to a quantized intrinsic localized mode (ILM).
The quantum ILM energy levels are then  observed  as peaks in the RRS spectrum
\cite{swanson}.
Measurements in  \cite{swanson} also found small magnitude 
ILM-plus-fundamental side-peaks. These  correspond to more complicated 
situations, when dynamics of at least two neighboring trimers are involved.
One of the trimers is excited   into an N-phonon state, and the other one 
is in a one-phonon state. The amplitude of the corresponding peaks 
is much less than the amplitude  of pure ILM peaks \cite{swanson}.
This fact is readily explained in the local picture of a RRS event 
described above. Dynamics of two trimers may become excited by 
multiple-hop events. 
These multiple-hop events 
have much less probability
than the event in  Fig.1, due to the very high degree of localization in 
$PtCl$ because of its very strong charge disproportionation 
\cite{swanson,bishop}.
An important issue is why, if two-trimer excitations do occur, they appear as
(N-1,1) modes, such that there are $N-1$ phonons on one trimer, and one phonon 
on the other. Other possible modes, for instance (N-2,2) modes are 
apparently not 
observed in the RRS spectrum within the experimental resolution 
\cite{swanson}. The answer to this puzzle  
seems to be in 
the bosonic nature of phonons. Bosons tend to bind together, or "condense",
into composite states.
The expression  for the probability amplitude for a boson to branch to some quantum state
has a multiplication factor $\sqrt{M+1}$, where $M$ is the number of bosons already
in this state. Therefore, if $N$ phonons are emitted by
an exciton and 
are allowed to branch into  
two states (in our case two neighboring trimers), the largest
probability is for all of them to go to the same trimer, and then 
the next most probable configuration is the (N-1,1) configuration.
We reiterate that the validity of the whole  picture described above 
is based upon the strongly localized nature \cite{swanson} of $PtCl$, i.e., 
upon the fact that being on 
a particular trimer is close to a well defined  quantum state. 
As  noted above, the energy of the $k$=$0$ optical phonon in 
$PtCl$ is $437$ K. Measurements \cite{swanson} were performed on samples
cooled to $12$ K. Even allowing for some local heating of the sample
by the laser light, it is impossible that the temperature of the
sample during the measurements becomes close to the thermal excitation threshold
of optical phonons. Therefore, 
there  cannot be thermalized optical phonons in the system.
In spite of that, measurements \cite{swanson} revealed clearly pronounced
"ILM-plus-fundamental" features. These features are anti-Stokes-like lines,
corresponding to a transformation of a pre-existing optical phonon into an
ILM by an absorption
 of a photon.
This means that the system  clearly has  pre-existing optical
phonons and is, therefore, not in  thermal equilibrium. 
 In order for such a non-equilibrium  phenomenon to be possible,
optical phonons should have a long lifetime,
namely the condition  $\tau_{phonon} G \sim 1$ should be satisfied, where $G$ is the number of 
photons absorbed 
in the sample per second per unit $PtCl$ cell. A detailed study of this 
interesting phenomenon
requires further experiments, particularly  of the 
anti-Stokes component of the RRS spectrum.  
It is worthwhile noting that non-thermal phonon distributions have
been invoked recently in other strongly coupled electron-phonon materials
(e.g. conjugated polymers and proteins \cite{vardeny,friedrich1,friedrich2}).
Also the long, multitimescale relaxation of
{\it classical} ILMs has been observed in numerical simulations
\cite{kim} on nonlinear lattices.

The usual adiabatic parameter, which is the ratio of characteristic phonon and
electron frequencies is small in PtCl. Therefore,  the 
amplitude of Peierls distortion should be well described in the adiabatic
approximation. On the other hand, as was shown in \cite{horowitz}, the 
purely adiabatic theory fails to describe dynamic optical absorption 
attributed to breathers. This is related to the fact that, although 
the adiabatic approximation may well describe the amplitude of the
wave function, it fails to correctly incorporate phase effects. 
The phase of the wave function does not play a role if one calculates
the amplitude of the Peierls distortion, since this calculation 
seeks the minimum of the adiabatic energy and is phase-insensitive. 
However, phase effects play a direct role in, e.g.,   the optical absorption,
where effects of constructive and destructive interference are 
present. 
One goal of our work here  was to check the limits of validity of the 
adiabatic approximation for the calculation of energies of quantum ILM's.
Our conclusion is that, although the adiabatic approximation 
can work well
for a calculation of the linear part of the energies, it may fail if one
calculates the nonlinear corrections, which correspond to the binding 
energies of phonons in  bound states. Since these binding energies are typically small,
 they may be strongly influenced by 
non-adiabatic effects.

A minimal phenomenological model
to describe  qualitative features of $PtCl$ is the two-band 
Peierls-Hubbard model, introduced for this purpose in \cite{bishop}. 
The ingredients of this model are the bond-length-dependent onsite energy difference between
$Pt$ and $Cl$ orbitals, the bond-length-dependent hopping matrix element between
$Pt$ and $Cl$ ions,  and  
the Hubbard repulsion $U$ for electrons on a $Pt$ atom. 
Lattice degrees-of-freedom are described
by the linear nearest-neighbor elastic spring constant $K$. 
Further studies \cite{batistic} have shown, that since the Coulomb interaction at the
length scale of one lattice constant  
is not fully screened, there is a need to take this interaction into  account,
assuming some reduced effective charges on neighboring sites. 
The Hamiltonian of our model is
\begin{eqnarray}
\label{4}
H = \sum_{l = - \infty, \sigma}^{\infty}
 - t_0(1 - \alpha \Delta_l) \left( c^{+}_{l,\sigma}c_{l +1,\sigma
} + h. c.\right) +  
U\sum_{l= - \infty }^{\infty} {   n_{2l,\uparrow} n_{2l,\downarrow}} 
+ 
\\
\nonumber
+\sum_{l= - \infty}^{\infty} 
{\left[\frac{K \Delta^2_l}{2} + \frac{P_l^2}{2m_l}\right]}
 +
\sum_{l= - \infty}^{\infty}
{
V_c \frac{(n_l-Z_l)(n_{l+1}-Z_{l+1})}{R_{l,l+1}}
} 
- 
\sum_{l= - \infty}^{\infty}
{
\left[- \epsilon + \beta (\Delta_{2l+1} +\Delta_{2l}) \right] 
n_{2l}.  
}
\\
\nonumber
\end{eqnarray}
\noindent
Here $c^{+}_i$,$c_i$ are electron creation and annihilation operators, $t_0$ is the $Pt \leftrightarrow Cl$ transition amplitude,
$\alpha$ and $\beta$ are  the electron-phonon couplings strengths, $\Delta_l$ is the bond
length change for the $l$-th bond, $U$ is the electron repulsion on
$Pt$ sites, $\epsilon$ is the energy difference between  $Cl$ and
$Pt$ sites, $K$ is the linear elastic constant, $m_l$ is the mass of the
$l$-th atom, 
$V_c$ is a phenomenological parameter related to the charge screening, and
$Z_l$ is the positive ion charge at the site $l$ \cite{batistic}.  
We assume $Pt$ sites to take even indices. The filling is three electrons
(or one hole) per unit cell.
Calculations were performed on 12-site chains, using numerical exact diagonalization  
and the method  of increments described in \cite{malek}.
The lattice was treated adiabatically. 
The coupling constants
were chosen to fit   experimental results for the magnitude of the 
Peierls
distortion, positions and magnitudes of $Pt^{4+} -> Pt^{2+}$ and $Pt^{4+} -> Cl$ optical
absorption peaks, and the  optical phonon frequency. Then the nonlinear 
adiabatic potential for $Cl$ was calculated by fixing the Peierls-distorted 
positions of all sites except
one $Cl$ atom,  and calculating the energy of the 
full system as a function of the position of this chosen $Cl$ atom, see Fig.2a.          
The displacement $X$ is measured
from the undistorted phase. Positive $X$ means a displacement in the 
direction of the $Pt^{2+}$ site.  The minimum of the potential is $X=-0.37$\AA,
which is the value of the  Peierls distortion.  
At $X=0.17$\AA $\;$ the calculated potential has a non-analytic point. 
At this point the first derivative of the potential with respect to 
$X$ has a jump.
Physically this non-analytic point corresponds
to a transfer of a hole from the $Pt^{4+}$ site to the closest $Pt^{2+}$
site. The mechanism of this transfer is as follows. As $X$ increases,
the  electron-phonon coupling leads to an energy increase at the $Pt^{4+}$ site.
Two holes  on the $Pt^{4+}$ site
experience the Hubbard repulsion. 
At $X= 0.17$\AA $\;$ the Hubbard repulsion dominates and one hole is transferred to
the closest $Pt^{2+}$ site. In this situation, the potential suddenly softens.
This effect is related to the potential formation of a local kink-antikink
pairs in the system, which is illustrated in  Fig. 2b. In this figure we
plot the adiabatic energy for {\it two} $Cl$ atoms symmetrically displaced.
We observe a local minimum in the adiabatic energy, which   should 
be attributed to the creation of a kink-antikink pair. 
Having the adiabatic potential, Fig.2a, 
nonlinear corrections to quantum  levels of a $Cl^{37}$ site oscillating in this 
potential  may be 
accurately calculated using perturbation theory for a quantum anharmonic oscillator \cite{flugge}.
First we performed our calculations without the Coulomb term and found the nonlinear
softening of the potential to be seriously underestimated. We were not able 
to reproduce the experimental values for the nonlinear corrections 
with any reasonable choice of model parameters. One can view this fact
as a confirmation of the importance of intra-chain Coulomb interactions in
$PtCl$ \cite{batistic}. We then added the Coulomb term and were able to fit the measured
values of nonlinear corrections by the following set of  model 
parameters:  
$t_0=0.75$ eV, $U=2.44$ eV,
$\beta=1.77$ eV/\AA, $\alpha = 3.55$ eV/\AA, 
$\epsilon = 0.90$ eV, $K=9.7$ eV/\AA$^2$,
$V_c = 8.18$ eV $\cdot$ \AA. 
Our value of the Hubbard $U$ is close to $U=2.0$ eV, obtained in \cite{wada} by
quantum chemical configuration interaction calculations on a
$Cl-Pt-Cl$ cluster. This correspondence is reasonable, since the Hubbard U
is  characteristic of a highly localized $d$ orbital, and should not change
much when going from an atomic cluster to a crystal. The value of the 
Coulomb repulsion coefficient $V_c$ has the same order of magnitude
as the value $13.3$ eV$\cdot$ \AA  $\;$ found in \cite{batistic}   
using a Hartree-Fock approach.
The adiabatic potential Fig.2a is quadratic near the equilibrium
point and may be written as $E_{ad} = F_1 dx^2/2+ F_2 dx^3/3 + F_3 dx^4/4$. 
The quantum levels for the $Cl$ atom moving in this potential well are  then given by the 
approximate expression \cite{flugge}

\begin{eqnarray}
E_n = \hbar \omega \left[ (n + 1/2) +  (\frac{\epsilon_1}{\hbar \omega}  )^2 A_n^{(2)}
+ \frac{\epsilon_1}{\hbar \omega} B_n^{(1)} - 
(\frac{\epsilon_1}{\hbar \omega})^2 B_n^{(2)} \right].
\end{eqnarray} Here 
$A_n^{(2)}=\frac{15}{4} (n^2 +n + \frac{11}{30})$, 
$B_n^{(1)}=\frac{3}{4} (2 n^2 +2 n +1 )$,  
$B_n^{(2)}=\frac{1}{8} (34 n^3 +51 n^2 +59 n +21 )$,
$\omega=\sqrt{F_1/2m}$, $l=\sqrt{\hbar/m \omega}$,  $\epsilon_1=l^3 F_2/3$,
$\epsilon_2=l^4 F_3 /4$. 
In  Fig.3 we have plotted the results for energy levels shifts, 
together with experimental RRS results  for $PtCl^{37}$ \cite{swanson}.
Here $N$ is the number of phonons in the bound state. For small
$N$ we find  good agreement with the experiment. At $N=5$ the 
amplitude of the shift becomes larger, than our estimates. 
This suggests to an additional softening nonlinearity in the system
at  sufficiently large amplitudes. 
It may be that this nonlinearity is related to 
nonadiabatic effects, which become more important as the amplitude 
of the ILM increases.  

To conclude, intrinsic localized modes (ILMs) in $PtCl$ give an example of 
nonlinear, thermally non-equilibrium excitations in  crystal lattices.
Similar phenomena, probably related to energy localization and non-thermalization,
have been experimentally observed, especially via ultrafast spectroscopy, 
in several electron-phonon coupled  systems. 
These systems include polymers \cite{vardeny}, glasses \cite{friedrich1}
and  biological systems such as proteins
\cite{friedrich2}. The 
unique advantage of PtCl and related MX materials is that one can
obtain very good crystals with a known structure and  controllable and tunable
nonlinearity strengths. Therefore, PtCl provides an ideal opportunity for making quantitive models and
testing theories of intrinsic nonlinear  energy localization
(as well as extrinsic energy localization \cite{love}).  
In this paper, 
a two-band Peierls-Hubbard model with screened  Coulomb interaction, 
in the adiabatic approximation,  
has been applied to PtCl. 
We have found that this  gives a good  qualitative account of the main phenomena
related to the existence of intrinsic localized modes in $PtCl$, at least for
small numbers of bound quanta.
We have also shown, that the screened Coulomb interaction is a necessary ingredient
within the adiabatic model to explain the quantitative magnitude
of nonlinear shifts in the resonant Raman spectrum.  
We have observed that there exists an additional source of  softening in
the system at sufficiently large amplitudes. 
We suspect this is  related to non-adiabatic effects;
future investigations will be focused on a quantitive theory of 
these effects.
Other important directions for  future experimental and theoretical research
include measuring  life-times of quantum ILMs, understanding
intrinsic  
mechanisms
for their decay and their interactions with impurities,    as well as 
studying photoexcited ILMs \cite{wang}.
PtCl offers the first controlled experimental possibility to  investigate these
fundamental questions with wide consequences for energy localization and
transport in strongly correlated hard, soft, and biological electronic 
materials.

We are grateful for stimulating discussions with S. Aubry, 
 A. Shreve,
B. Swanson  and C. R. Willis.  Work at Los Alamos is supported by the U.S.D.o.E, under contract
W-7405-ENG-36.

\noindent
\newpage
FIGURE CAPTIONS
\\
\\
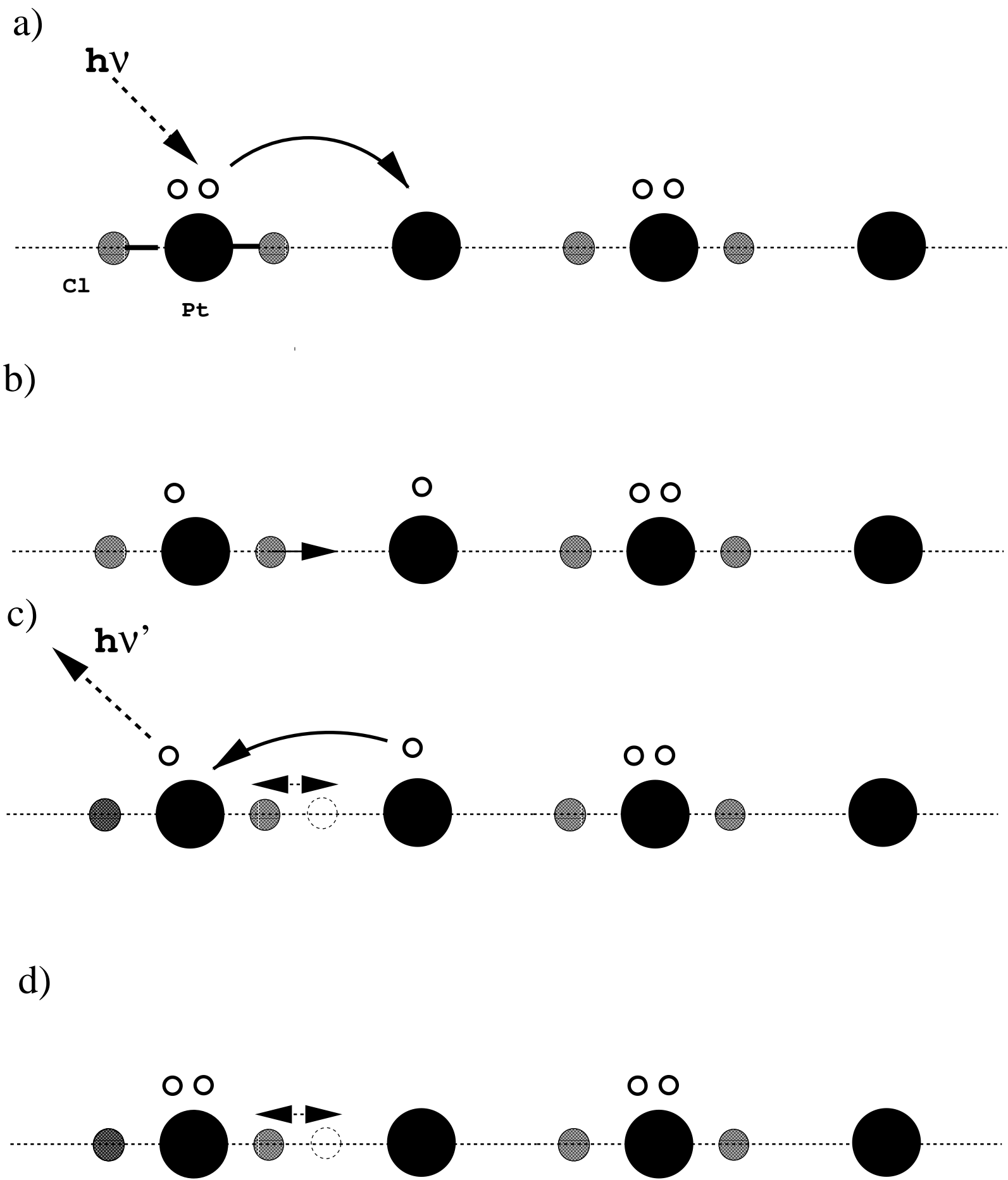
Fig.1: 
A simple picture of a resonant Raman scattering event in the localized (atomic) limit.
\\
\\
Fig.2:
(a) Adiabatic potential for a single $Cl$ atom. The zero of  energy corresponds
to the equilibrium position in the Peierls-distorted phase. 
$X$ is the shift of the $Cl$ atom, measured from the 
undistorted phase. 
(b)
Adiabatic potential for two symmetrically displaced
$Cl$ atoms in one trimer.
\\
\\
Fig.3:
The absolute value of the nonlinear energy level shift. 
Dots correspond to the theoretical
prediction. Circles are the experimental results for $PtCl^{37}$\cite{swanson}; the size of a circle
gives the experimental uncertainty due to a finite RRS peak width.
Lines are guides to the eye.

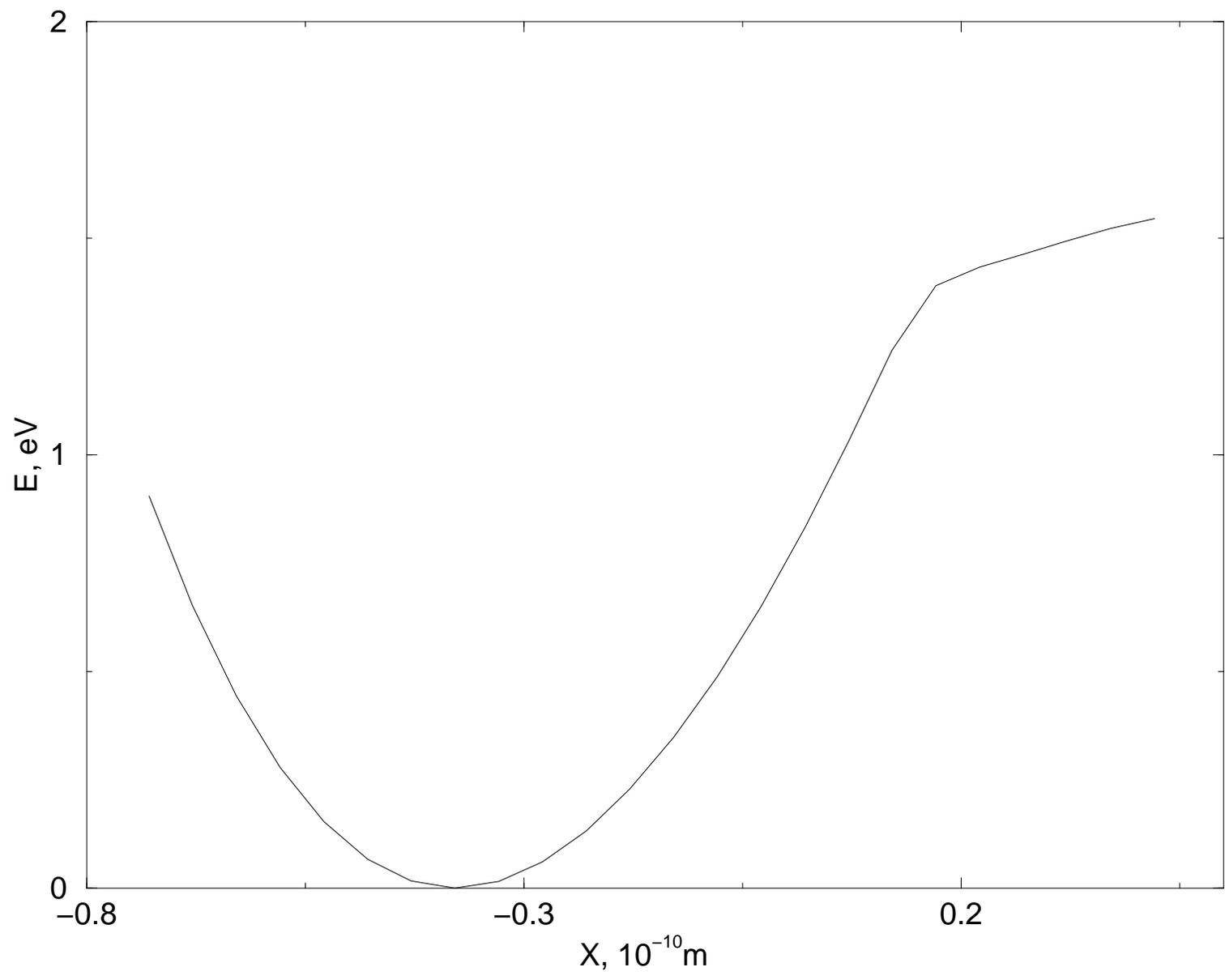
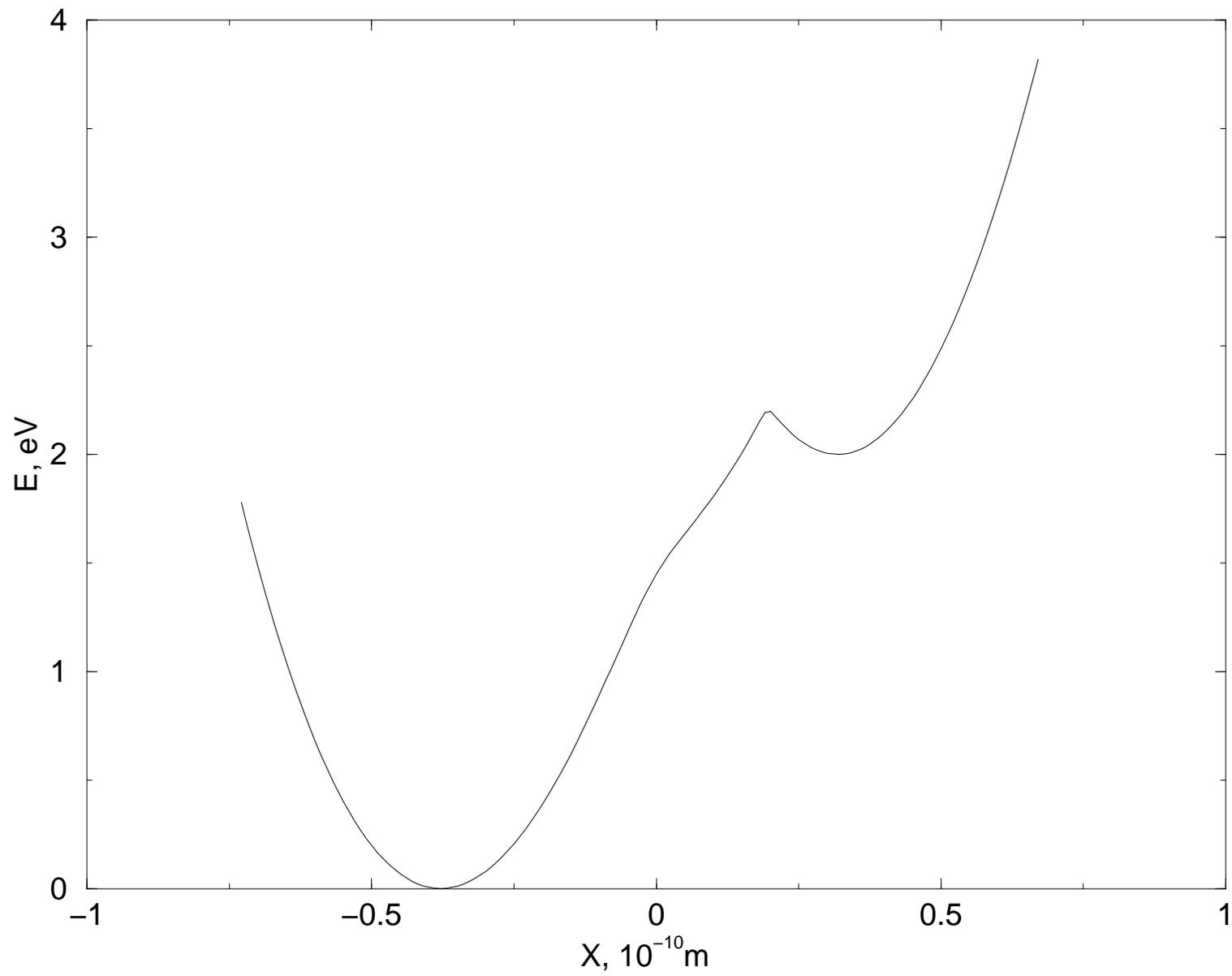
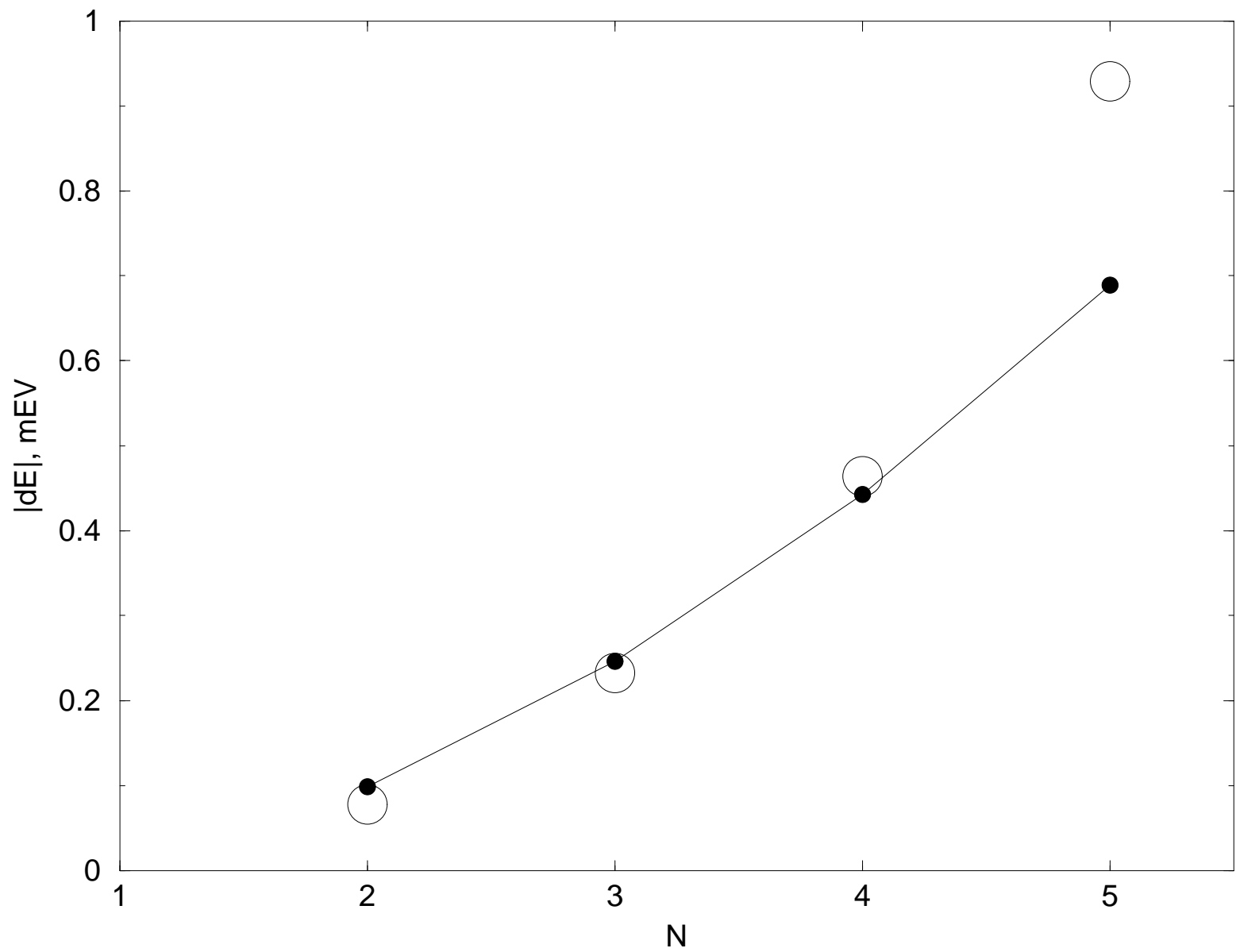
\end{document}